\documentclass[12pt]{iopart}
\usepackage{graphicx}

\begin{document}

\title{ Density Anomaly in Core-Softened Lattice Gas  } 

\author{Aline Lopes Balladares}
  
\address{Instituto de F\'{\i}sica, Universidade Federal do Rio
  Grande do Sul \break Caixa Postal 15051, 91501-970, Porto Alegre,RS, Brazil}

\author{Marcia C. Barbosa\footnote[3] {To whom correspondence should be addressed (barbosa@if.ufrgs.br)}}

\address{Instituto de F\'{\i}sica, Universidade Federal do Rio
  Grande do Sul \break Caixa Postal 15051, 91501-970, Porto Alegre,RS, Brazil}

\begin{abstract}
A two dimensional lattice gas model with ''core-softened'' potential
is investigated. Two liquid phases and density anomaly are found.
The demixing phase transition between the two liquid phases end 
at a tricritical point that is also the terminus of a critical line.
The density anomaly is shown to be related to this continuous line.
 \end{abstract}

~

\maketitle
\section{Introduction}

The phase behavior of single component systems as particles 
interacting via the so-called core-softened $(CS)$ potentials
is receiving a lot of attention recently. 
These potentials exhibit a repulsive core with
a softening region with a shoulder  or 
a ramp \cite{Ja99}-\cite{Wi02}.
These  models originates
 from the disere of constructing a 
simple two-body isotropic potential capable of describing
the complicated features of systems interacting via anisotropic 
potentials. This 
procedure generates models that are analytically and
computationally tractable and that one hopes are capable
to retain the qualitative features of the real complex systems.

The physical motivation behind these studies is the recently acknowledged 
possibility  that some single 
component systems  display coexistence
between two different liquid phases \cite{Fr01}\cite{Po92}-\cite{Gl99}. This 
opened the 
discussion about the relation between presence of 
two liquid phases, the existence of thermodynamic anomalies in liquids
and the form of the potential.
The case of water is probably the most intensively studied.  For
instance, liquid water has a maximum as  a  function of temperature in 
both density and compressibility \cite{De96}.
 It has been proposed some time ago that
these  anomalies might be associated with a critical point at the terminus of
a liquid-liquid line, in the unstable supercooled liquid region \cite{Po92},
at high pressures. This hypothesis has been supported
by a  varied experimental data \cite{Mi98}\cite{angell} that show 
 that thermodynamic singularities  are present in supercooled water, around
\( 228K \) and at atmospheric pressure. In spite of the limit of \( 235K \)
below which water cannot be found in the liquid phase without crystallization,
two amorphous phases were observed at much lower temperatures \cite{Mi84}.
There is evidence, although yet under test, that these two amorphous phases
are related to two liquid phases in  fluid water \cite{Sm99}\cite{Mi02}.

Water is not an isolated case. There are also other examples of  tetrahedrally
bonded molecular liquids such as phosphorus \cite{Ka00}\cite{Mo03} 
and amorphous  silica \cite{La00} that also are good candidates for having 
two liquid phases. Moreover, other materials such as liquid metals 
\cite{Cu81} and graphite \cite{To97} also exhibit thermodynamic anomalies. 

Acknowledging that CS potentials may engender a demixing
transition between two liquids of different densities, a number of 
CS potentials were proposed to model the anisotropic systems described above.
The first suggestion was made many years ago by Stell and coworkers 
in order to explain the isostructural solid-solid
transition ending in a critical point \cite{He70}-\cite{Ho73}.
Debenedetti et al. \cite{De91}  using general thermodynamic 
arguments, confirmed 
that a CS can lead to a coefficient of thermal expansion negative and
consequently to density anomaly. This together
with the increase of the thermal compressibility has
been used as indications of the presence of two liquid
phases \cite{St98}\cite{St00} which may be hidden beyond the homogeneous
nucleation.  The difficulty  with these approaches is 
that continuous potentials usually lead to crystallization 
at the region where the critical point would be expected.

In order to circumvent this problem, we study the
effect of CS potentials in a lattice. Even
thought the lattice is not realistic, it allows
us to explore the phase space  in a easier  way. 
In this work 
we analyze a two dimensions lattice gas  with 
nearest-neighbors repulsive interactions and next-nearest-neighbors
attraction. The system is in contact with a reservoir of 
temperature and particles. We show that this very simple
system exhibits  both density anomaly and two liquid phases.
However, instead of 
having a critical point ending the coexistence line between
the two liquid phases as one usually would expect, it has a tricritical point.
The connection between the presence of criticality and the density
anomaly is also shown. 

The reminder of the paper goes as follows. 
In sec. 2 the model is presented and the zero temperature
phases are introduced, the mean field analysis is shown on
sec. 3, results from simulations are discussed in
sec. 4 and our findings are summarized in sec. 5.

\section{The model and its  ground state}

Our system consists of a two-dimensional
square lattice with $N$ sites.  Associated
to each site there is an occupational variable, $\sigma_i$. If the  site 
is occupied by a molecule,  $\sigma_{i}=1$, otherwise
$\sigma_{i}=0$.
Each site interacts with its nearest neighbors with repulsive interactions
and with its next-nearest neighbors with attractive  interactions 
( see Fig.1). Therefore
the Hamiltonian  of this system is given by
\begin{eqnarray}
H=-V_{1}\sum_{<i,j>}\sigma_{i} \sigma_{j} -V_{2}
\sum_{<<i,j>>} \sigma_{i}\sigma_{j}
\label{H}
\end{eqnarray}
where, $<i,j>$ represents the sum over 
the  nearest
neighbors  and $<<i,i>>$ is the sum over the next nearest
neighbors.
Our system is in contact with a temperature and particle  reservoirs.
 The grand potential  is given by:
\begin{eqnarray}
\Phi=\langle \mathcal{H} \rangle -TS
\label{phi}
\end{eqnarray}
where $\mathcal{H}$ contains the internal energy and the
contribution due to the chemical potential $\mu$, namely
\begin{eqnarray}
\mathcal{H}=H-\mu\sum_{i}^{N} \sigma_i \; .
\label{calH}
\end{eqnarray}

Let us now consider the ground state properties of this
model. The Hamiltonian Eq.~(\ref{calH}) allows for 
a number of different configurations, however, due to 
the lattice symmetry and the nature of  interaction, just five of them 
might exhibit lower energy as the chemical potential is varied. They are ( see Fig.(2))

\begin{itemize}
    \item Dense liquid (dl):
    \begin{eqnarray}
\phi_{dl}=\frac{\Phi_{dl}}{N}=-2V_{1}-2V_{2}-{\mu}   \; .
\label{DL}
\end{eqnarray}

\item Uniformly diluted liquid (udl) :

    \begin{eqnarray}
\phi_{udl}=\frac{\Phi_{udl}}{N}=-V_2-\frac{1}{2}{\mu} \; .
\label{udl}
\end{eqnarray}
    
\item Structured diluted liquid (sdl):

    \begin{eqnarray}
\phi_{sdl}=\frac{\Phi_{sdl}}{N}=-\frac{1}{2}V_1-\frac{1}{2}{\mu}  \;.
\label{sdl}
\end{eqnarray}
    
\item Semi-diluted liquid (semi-dl):
   
 \begin{eqnarray}
\phi_{semi-dl}=\frac{\Phi_{semi-dl}}{N}=-V_{1}-V_{2}-\frac{3}{4}{\mu}  \;.
\label{semi-dl}
\end{eqnarray}
    
\item Gas (gas):
    
\begin{eqnarray}
\phi_{gas}=\frac{\Phi_{gas}}{N}=0  \; .
\label{gas}
\end{eqnarray}
\end{itemize}
Here $\phi$ is the grand potential per site. 

Comparing these expressions for different chemical potentials leads
to the following zero temperature phase-diagram.
For positive  chemical potential,  $\mu>>|V_1|$ and $\mu>>V_2$, the
lower grand potential is associated with the dense liquid phase.
As the chemical potential is reduced, the interactions
between molecules become relevant. The first-neighbors repulsion
together with the second-neighbors attraction favors the formation
of the udl phase. At  $\mu=-4V_{1}-2V_{2}$ there is 
a phase  transition  between the dense
liquid phase and the  uniformly  diluted liquid phase. If the 
chemical potential is decreased even further,   at $\mu=-2V_2$ there
is a transition between the uniformly diluted liquid phase and the  gas phase.

\begin{figure}
\begin{center}
\includegraphics[height=7cm,width=7cm]{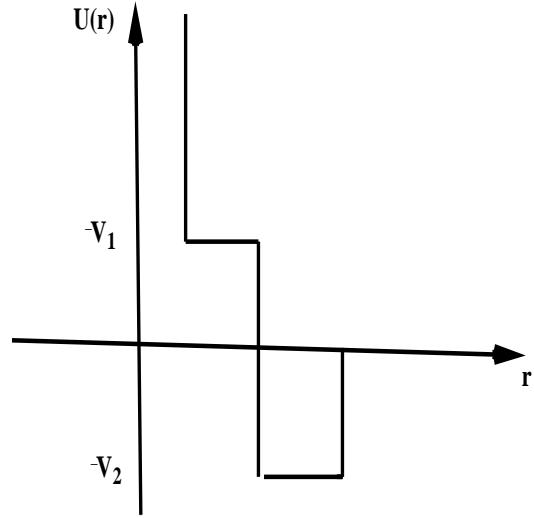}
\caption{Schematic form of the interaction potential.}
\end{center}
\end{figure}
\begin{figure}
\begin{center}
\includegraphics[width=3cm,height=3cm,angle=-90]{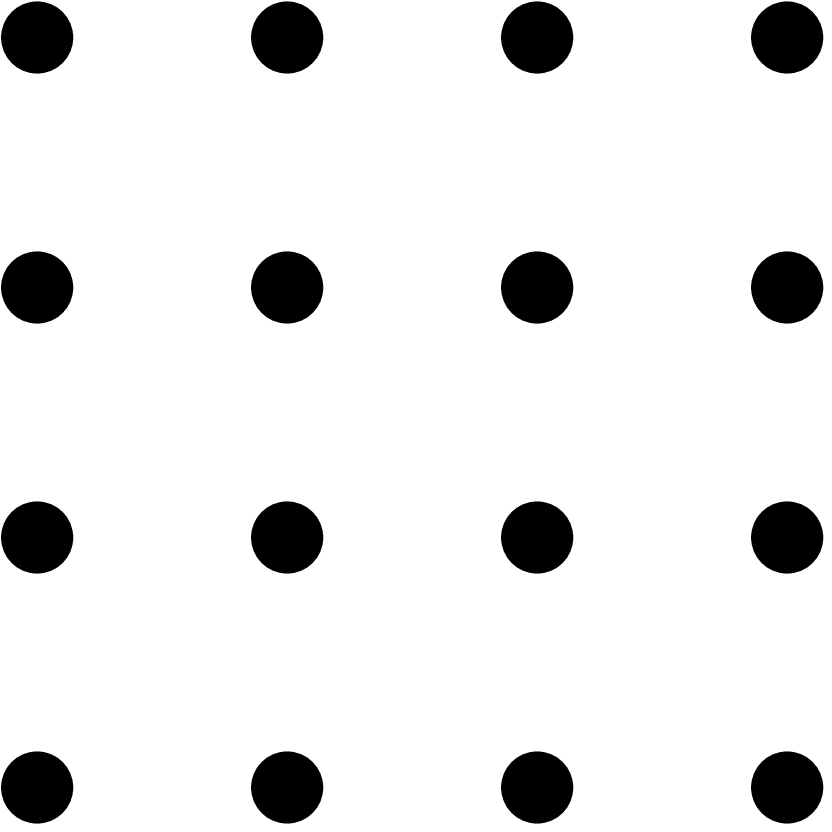}
\hspace{1cm}
\includegraphics[width=3cm,angle=-90]{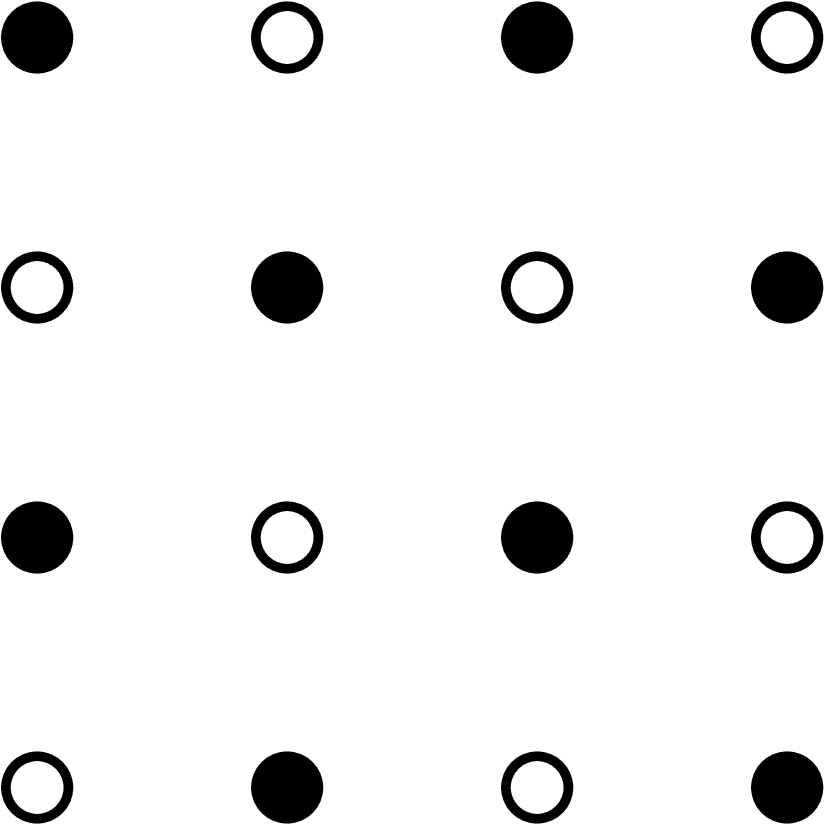}

($a$) \hspace{4cm} ($b$)

\includegraphics[width=3cm,angle=-90]{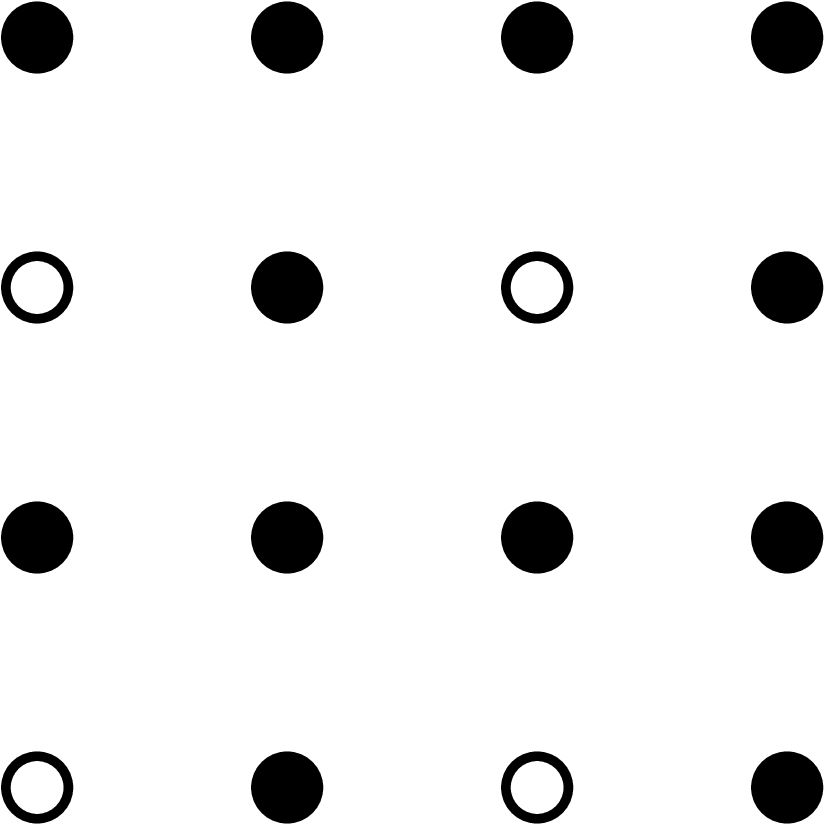}
\hspace{1cm}
\includegraphics[width=3cm,angle=-90]{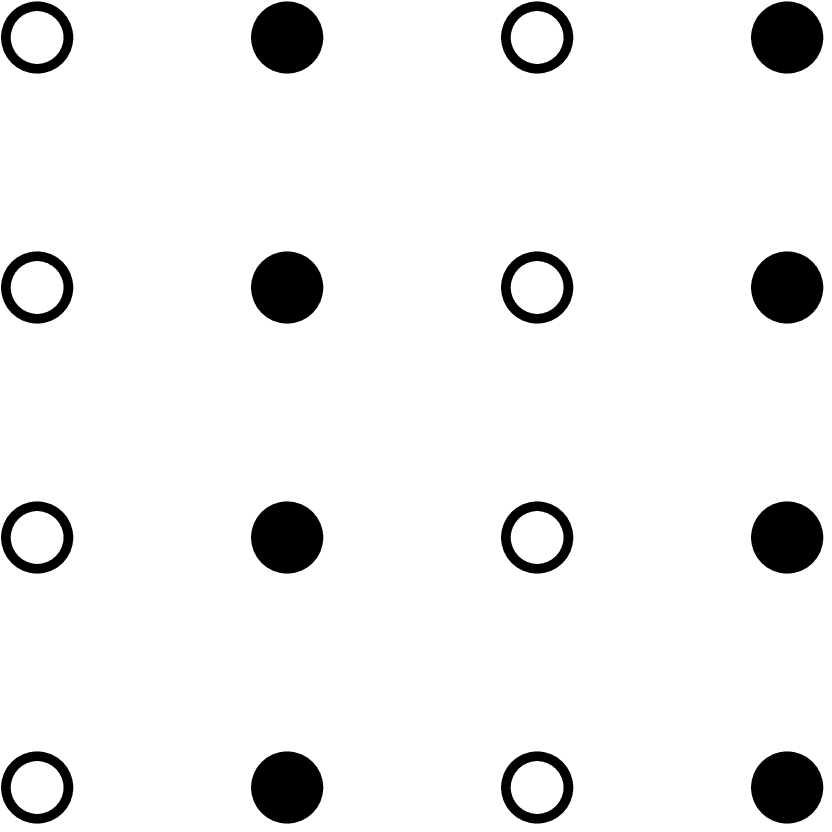}

($c$) \hspace{4cm} ($d$)

\includegraphics[width=3cm,angle=-90]{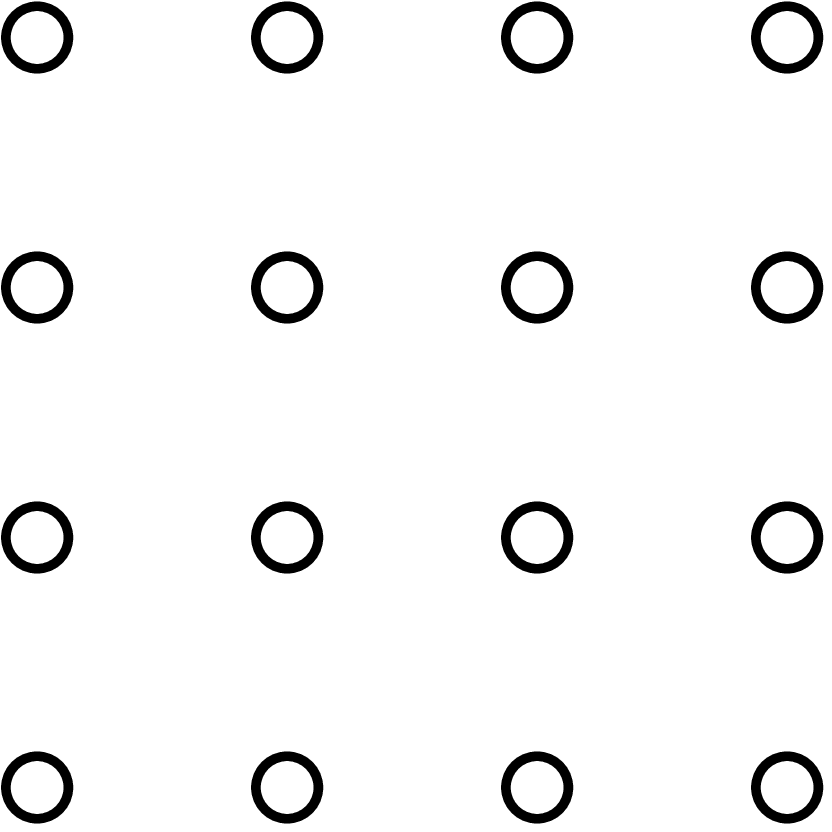}
\hspace{1cm}
\includegraphics[width=3cm,angle=-90]{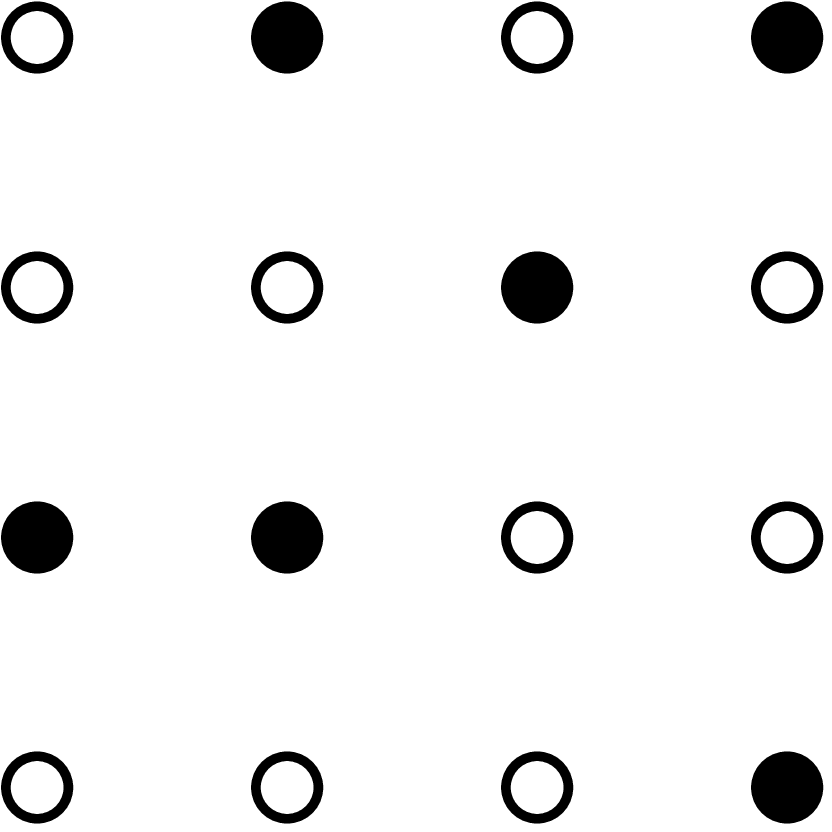}

($e$) \hspace{4cm} ($f$)
\caption{ $(a)$ dense liquid phase, $(b)$ uniformly dense liquid phase,
 $(c)$ semi-diluted liquid phase, $(d)$ structured diluted liquid phase,
$e$ gas and $(f)$ fluid phase.}

\end{center}

\end{figure}
\begin{figure}
\begin{center}
\includegraphics[height=2cm,width=2cm]{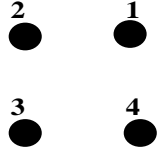}
\caption{The sub-lattices.}
\end{center}
\end{figure}

\section{Mean-Field}

Now, let us exam the phase-diagram for nonzero temperatures 
employing a mean-field approximation.
The symmetry of the different phases can be better
visualized if the square  lattice is divided into four sub-lattices
as illustrated in Fig.(3).  In this case, the density 
of the sub-lattice $\alpha$ is given by:
\begin{eqnarray}
\rho_\alpha=\frac{4}{N}\sum_{j\in\alpha}^{}\sigma_j
\label{sub}
\end{eqnarray}
where the sum  $j\in \alpha$ is over one of  the sub-lattices $\alpha=1,2,3,4$. Note that  the density of each
sub-lattice varies between 0 and 1. Using the sub-lattice division, 
 the Hamiltonian given for the Eq.~(\ref{calH}) can be written as:
\begin{eqnarray}
{\mathcal{H}}=-\sum_{\alpha=1}^4\sum_{i\in\alpha}\mu_{\alpha}^{\emph{eff}}
\left(\{\sigma_{i}\}\right)\sigma_{i}
\label{subcalH}
\end{eqnarray}
where
\begin{eqnarray}
\mu_{\alpha}^{\emph{eff}}\left(\{\sigma_{i}\}\right)=
\mu_{\alpha}+\sum_{\beta=1}^4\sum_{j\in\beta}J_{ij}\sigma_{j}
\label{mueff1}
\end{eqnarray}
is  the  effective chemical potential acting in an ideal 
sub-lattice $\alpha$. Here  
\begin{eqnarray}
\mu=\frac{1}{4}\sum_{\alpha=1}^4 \mu_{\alpha}
\label{mu}
\end{eqnarray}
is the chemical potential contribution due to the particles 
reservoir, while the second contribution in Eq.~(\ref{mueff1}) is
 due to the interaction with  other sub-lattices. The mean-field
approximation we employ is to take  the average
of this last term, namely:
\begin{eqnarray}
\bar{\mu}_{\alpha}^{\emph{eff}}\left(\{\sigma_{i}\}\right)=
\mu_{\alpha}+\sum_{\beta=1}^{4}\sum_{i\in\beta}J_{ij}\langle\sigma_{j}\rangle
=\mu_{\alpha}+\sum_{\beta=1}^{4}\epsilon_{\alpha\beta}\rho_{\beta},
\label{mueff2}
\end{eqnarray}
where 
\begin{eqnarray}
\rho_{\beta}=\frac{4}{N}\sum_{j\in\beta}\langle\sigma_{j}\rangle
\label{rho_alfa1}
\end{eqnarray}
is the  density of the sub-lattice $\beta$ and where 
\begin{equation}
\epsilon_{\alpha\beta}=\sum_{j(\neq i)}J_{ij} ,\quad
i\in\alpha,\quad j\in\beta
\label{epsilon}
\end{equation}
is  an interaction parameter.
The mean-field Hamiltonian then becomes
\begin{eqnarray}
{\mathcal{H}}^{mf}=-\sum_{\alpha=1}^{4}\sum_{i\in\alpha}\left(\sum_{\beta=1}^{4}\epsilon_{\alpha\beta}
\rho_{\beta}+\mu_{\alpha}\right)\sigma_{i}-\frac{1}{2}\sum_{\alpha=1}^{4}\frac{N}{4}
\sum_{\beta=1}^{4}\epsilon_{\alpha\beta}\rho_{\alpha}\rho_{\beta}
\label{Hmf}
\end{eqnarray}
where the second term corrects for over-counting. It is
straightforward to show that using Eq.~(\ref{Hmf}) 
 the mean field approximation 
for the grand potential per site is given
by
\begin{eqnarray}
\phi^{mf}& = &
-k_{B}T\ln{2}-\frac{k_{B}{T}}{4}\ln{\cosh{\left[-\frac{\beta}{2}
\sum_{\beta=1}^{4}\left(\epsilon_{\alpha\beta}\rho_{\beta}+
\mu_{\alpha}\right)\right]}} \nonumber\\
& & -\frac{1}{8}\sum_{\alpha=1}^{4}\left(\sum_{\beta=1}^{4}
\epsilon_{\alpha\beta}\rho_{\beta}+\mu_{\alpha}\right)-\frac{1}{8}
\sum_{\alpha=1}^{4}\sum_{\beta=1}^{4}\epsilon_{\alpha\beta}\rho_{\alpha}\rho_{\beta}\; .
\label{phimf}
\end{eqnarray}

The sub-lattice density can be derived both 
from Eq.~(\ref{rho_alfa1}) and from the mean field
 grand potential by deriving it with respect to the
chemical potential, namely
\begin{eqnarray}
\rho_{\alpha}=-4\left(\frac{\partial\phi^{mf}}{\partial\mu_{\alpha}}
\right)_{T,\mu_{\alpha\neq\beta}}
\quad\alpha=1,\ldots,4 \;\; .
\label{rho_alfa2}
\end{eqnarray}
In both cases, the derivation leads to;
\begin{eqnarray}
\rho_{\alpha}=-\frac{1}{2}\tanh{\left[\frac{\beta}{2}\sum_{\beta=1}^{4}
\left(\epsilon_{\alpha\beta}\rho_{\beta}+\mu_{\alpha}\right)\right]}-\frac{1}{2},
\quad \alpha=1,2,3,4 \;\;.
\label{rho_alfa3}
\end{eqnarray}

The phase-diagram, illustrated in Fig.(4), is
obtained from solving   Eq.~(\ref{rho_alfa3}) and Eq.~(\ref{phimf}) for
different temperatures and chemical potentials. Here $V_1=-1$ and 
$V_2=1$ are fixed. Different choices of the interaction parameters
do not change the phase-diagram qualitatively.
 At high temperatures, all
the sub-lattice  densities  are   $\rho_\alpha=1/2$ and the system
is in the fluid phase. As the temperature is decreased at high
chemical potential, $\bar{\mu}=\mu/|V_1|>2$,  the sub-lattices become
full and the system goes from the fluid to the dense liquid phase. 
For  negative 
chemical potentials, $\bar{\mu}<-2$, the sub-lattices get  empty, 
$\rho_{\alpha}=0$,  and  the system goes from 
the fluid to the   gas 
phase. Between these  two limits, $2>\bar{\mu}>-2$, as
the temperature is decreased  two 
opposite sub-lattices become
empty while the other two get full. The system goes
from the fluid phase to the udl phase through a continuous phase 
transition, $\bar{T}_c(\bar{\mu})$.
The low temperature coexistence lines
between the udl phase and the dl phase and between the udl and the gas phase
are  obtained by comparing the  grand potentials per particle, Eq.~(\ref{phimf}), of these phases. The udl-dl phase boundary  at $\bar{\mu}=2$
and the udl-gas coexistence line at  $\bar{\mu}=-2$ meet the 
critical line   $T_c(\bar{\mu})$ at two tricritical points at
$(\bar{T}_t=0.924,\bar{\mu}_t=2)$ and $(\bar{T}_t=0.924,\bar{\mu}_t=-2)$ respectively.

\vspace{1.5cm}

\begin{figure}
\begin{center}
\includegraphics[height=7cm,width=7cm]{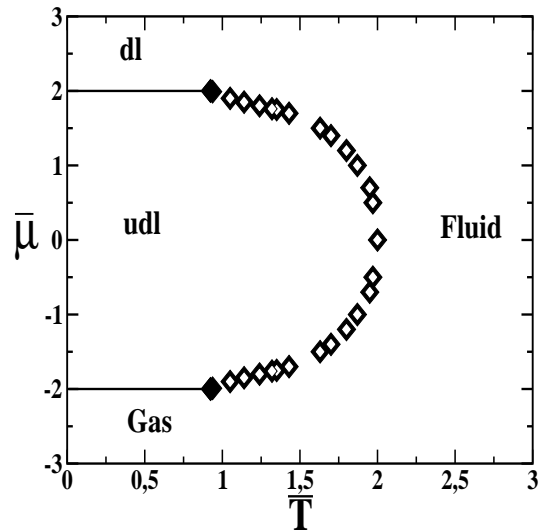}
\caption{Phase Diagram for $V_{2}/|V_{1}|=1$. The empty  diamonds are continuous lines, the solid lines are  first order transitions  and the filled diamonds are the  tricritical points.}
\end{center}
\end{figure}

\section{Monte Carlo Simulation}

The rather simple mean field approach employed in the previous 
session is unable to account for the density anomaly. For
investigating the possibility of a density anomaly in
our  potential,  Monte Carlo simulations in
grand canonical ensemble were performed. 
The Metropolis algorithm was used to study square 
$LxL$ lattice and $V_2/|V_1|=1$.  Different 
system sizes $L=10,20,30,50$ were investigated. Equilibration
time was $1 000 000 $ Monte Carlo time steps for each lattice site. 

The  Monte Carlo simulations at fixed  chemical
potentials  give the following results.
At high temperatures, the density of each sub-lattice 
is around  $\rho_\alpha=1/2$ with sites randomly occupied in
each sub-lattice, so the system is in the 
fluid phase. As the temperature is decreased at a
fixed chemical potential,  $\bar{\mu}>2$, the density of
each sub-lattice increases continuously to one. The system
goes from the fluid phase to the dense liquid phase with
no phase transition. 
If the system is cooled from the fluid phase 
at low chemical   potential, 
 $\bar{\mu}<-2$, the density decreases continuously from the 
fluid phase to the gas phase with no phase transition.
As the  temperature  is decreased at fixed chemical
potential,    $-2<\bar{\mu}<2$,  one
finds that two opposite sub-lattices ( $1$ and $3$ for example) become
full while the other two sub-lattices ( $2$ and $4$) get empty what
characterizes the udl phase. Fig.(5) shows that at
$\bar{\mu}=1.2$ the density of each sub-lattice jumps
from the fluid phase value to the udl phase value at 
the critical temperature
$\bar{T}_c(\bar{\mu})\equiv T_c/|V_1|=1.18$ 
Simulations for various fixed chemical potentials, allow us to
find the  critical line $\bar{T}_c(\bar{\mu})$.

\vspace{1cm}

\begin{figure}
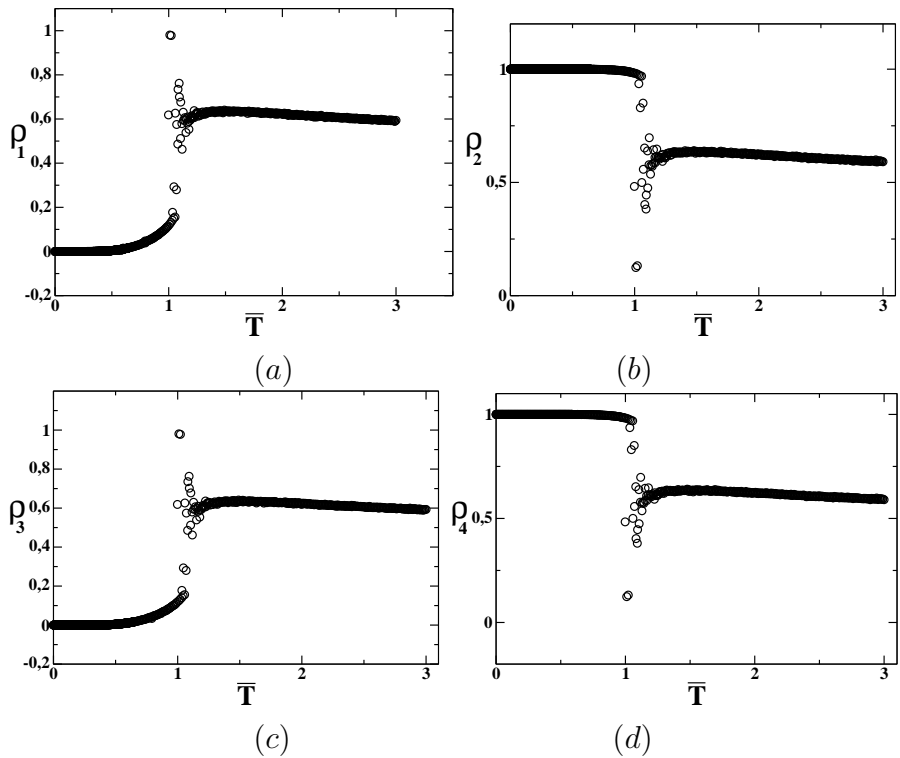

\begin{center}
\includegraphics[ clip, scale=0.24]{fig5a.eps}
\includegraphics[ clip, scale=0.24]{fig5b.eps}

($a$) \hspace{4cm} ($b$)

\includegraphics[ clip, scale=0.24]{fig5c.eps}
\includegraphics[ clip, scale=0.24]{fig5d.eps}

($c$) \hspace{4cm} ($d$)

\caption{Densities of  the $(a)$ sub-lattice $1$, $(b)$ sub-lattice $2$,
 $(c)$ sub-lattice $3$ and  $(d)$ sub-lattice $4$ for the 
lattice $20x20$ at $\bar{\mu}=1.2$. }
\end{center}
\end{figure}
In principle, 
the fluctuations observed in Fig.(5) suggest that
the transition at $\bar{T}_c(\bar{\mu})$ is continuous.
 This assumption is   supported by the 
increase in the specific heat at a fixed chemical
potential. Fig.(6) illustrates this increase at  $\bar{T}_c(\bar{\mu})=1.18$ 
for $\bar{\mu}=1.2$.

\begin{figure}
\begin{center}
\includegraphics[clip, scale=0.4]{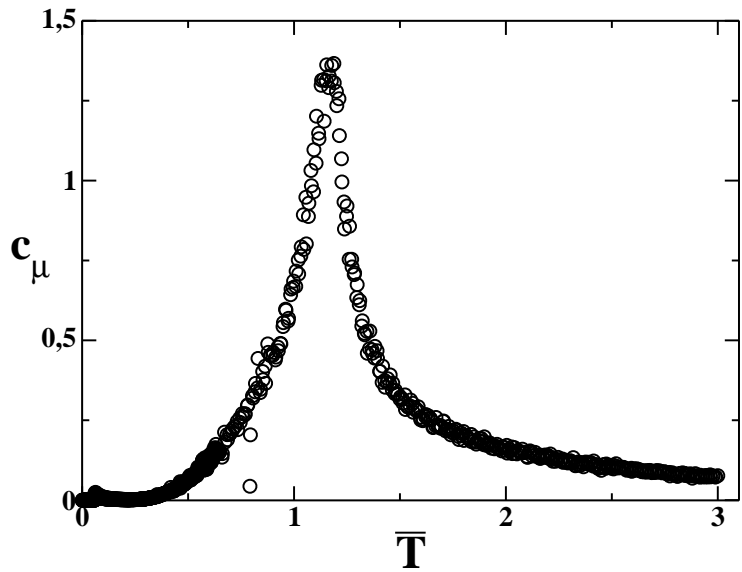}
\end{center}
\caption{Specific heat for the lattice $20x20$  and $\bar{\mu}=1.2$. }
\end{figure}

The presence of a minimum in the fourth-order 
Binder cumulant given by \cite{Ch86}:
\begin{eqnarray}
V_{L}=\frac{1-\langle\mathcal{H}^{4}\rangle}{3\langle\mathcal{H}^{2}\rangle^{2}}  
\label{binder}
\end{eqnarray}
is an indication of criticality. 
Fig.(7) shows $V_L$  for $\bar{\mu}=1.2$ obtained 
from  our simulation.  The 
minimum at $\bar{T}_c(\bar{\mu})=1.1787$ is a sign of 
 the presence of criticality.
The hypothesis that the  transition at $\bar{T}_c(\bar{\mu})$ 
is  first-order is eliminated by computing  the energy histogram shown in
Fig.(8). If the transition would be first-order,
two distinct peaks should be present.

\begin{figure}
\begin{center}
\includegraphics[clip, scale=0.4]{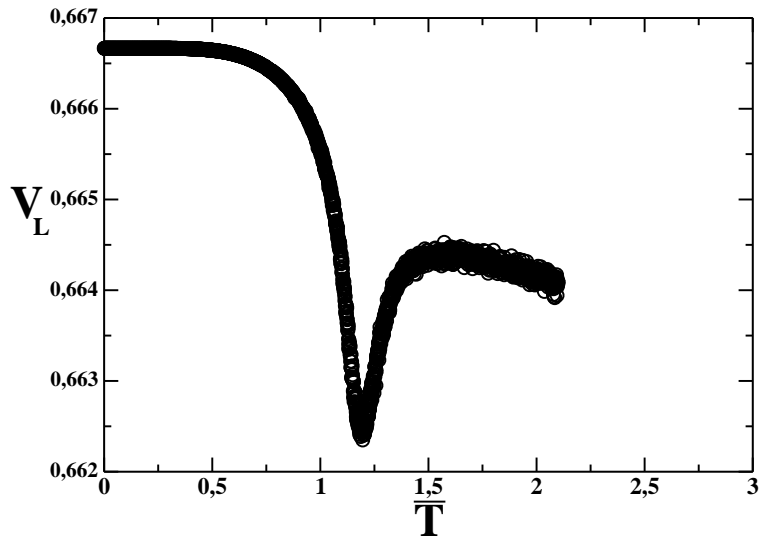}
\caption{Fourth-order Binder's  cumulant for  the lattice $20x20$ and
for $\bar{\mu}=1.2$.} 
\end{center}
\end{figure}

\begin{figure}
\begin{center}
  \includegraphics[clip, scale=0.4]{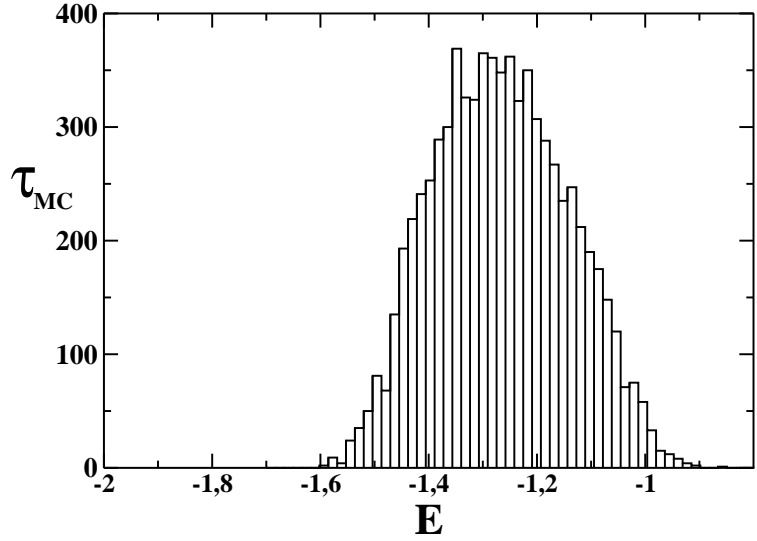}
  \caption{Energy versus Monte Carlo steps, $\tau_{MC}$ for the lattice $20x20$, $\bar{\mu}=1.2$ and $\bar{T}=1.18$.
  The presence of only one peak characterizes a continuous  transition.}
\end{center}
\end{figure}

In order to check the location
of the coexistence lines  
observed by the mean-field analysis,  simulations varying the 
chemical potential at
a fixed temperature  were performed. The results are shown in 
Fig.(9). 
For temperatures within the interval $0<\bar{T}<0.5$, as the 
chemical potential is increased the system exhibits two first-order
phase transitions, one between the gas and udl phase and 
another between the udl and the dense liquid  phase.

\begin{figure}
\begin{center}
\includegraphics[clip, scale=0.4]{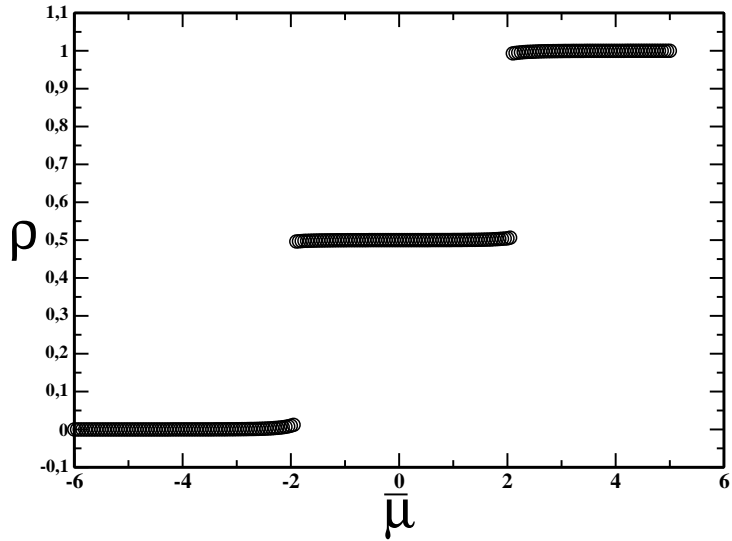}
  \caption{$\rho.vs.\bar{\mu}$: the first-order transitions
between the gas-udl and udl-dl    are illustrated.}
\end{center}
\end{figure}

The total density of the system at fixed chemical
potential is illustrated in Fig.(10). Differently
from the mean-field result, simulations show an anomalous behavior of
the density, instead of being monotonic
with temperature, the density at positive chemical potential increases
as the temperature is decreased, it has 
a maximum at a temperature of maximum density ($\bar{T}_{TMD}(\bar{\mu})$) and
then decreases. For negative chemical potentials, the complementary effect
is shown in Fig.(11), defining a temperature of minimum density ($\bar{T}_{TmD}(\bar{\mu})$).

\begin{figure}
\begin{center}
\includegraphics[clip, scale=0.4]{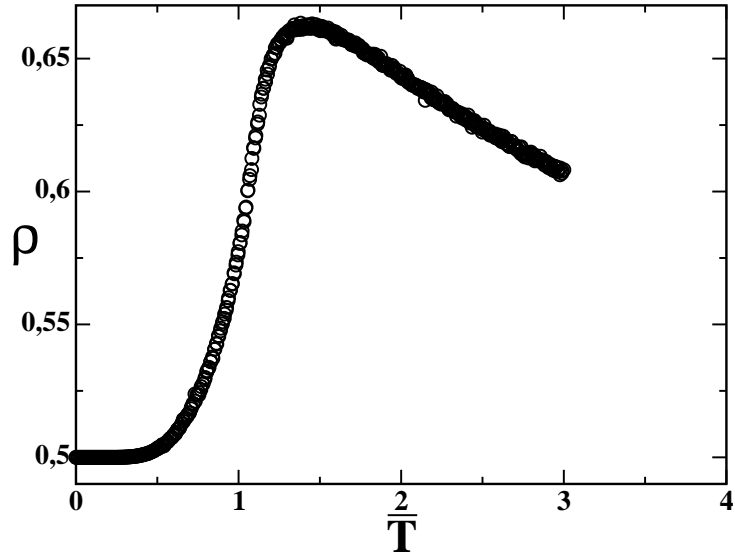}
  \caption{$\rho.vs.T$: the maximum in the density for  the lattice $20x20$ and
  $\bar{\mu}=1.4$ is shown.}
\end{center}
\end{figure}

\begin{figure}
\begin{center}
\includegraphics[clip, scale=0.4]{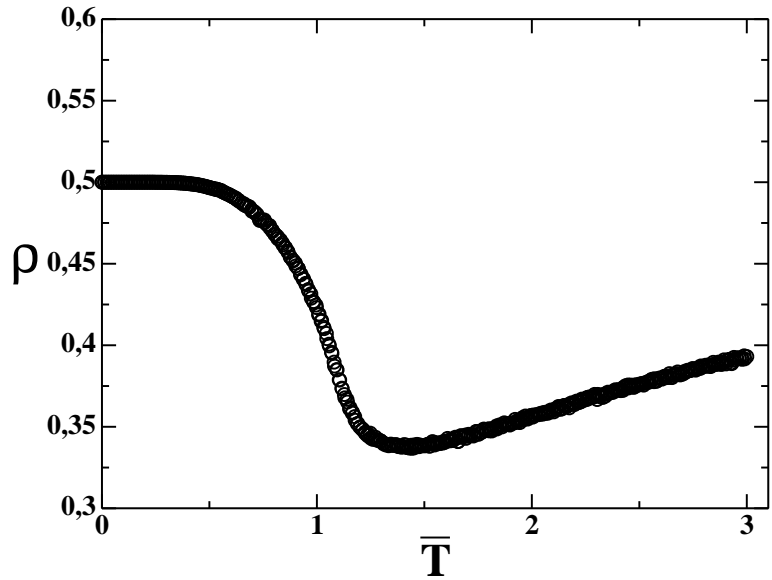}
  \caption{$\rho.vs.T$: the minimum in the density for the 
lattice $20x20$ and chemical  potential $\bar{\mu}=-1.4$ is shown.}
\end{center}
\end{figure}

Summarizing the results discussed above, the $\mu.vs.T$ phase-diagram is shown
in Fig.(12).   The critical line that separates
the fluid from the uniformly diluted phase joint the 
phase boundaries between  the uniformly diluted phase
and the dense liquid and gas phases at symmetric tricritical points at
 ($\bar{\mu}=2,\bar{T}=0.5237$) and  ($\bar{\mu}=-2,\bar{T}=0.5237$) 
respectively. The lines of temperature of maximum density  and minimum 
density at 
constant chemical potential  are
shown.

\begin{figure}
\begin{center}
\includegraphics[clip, scale=0.4]{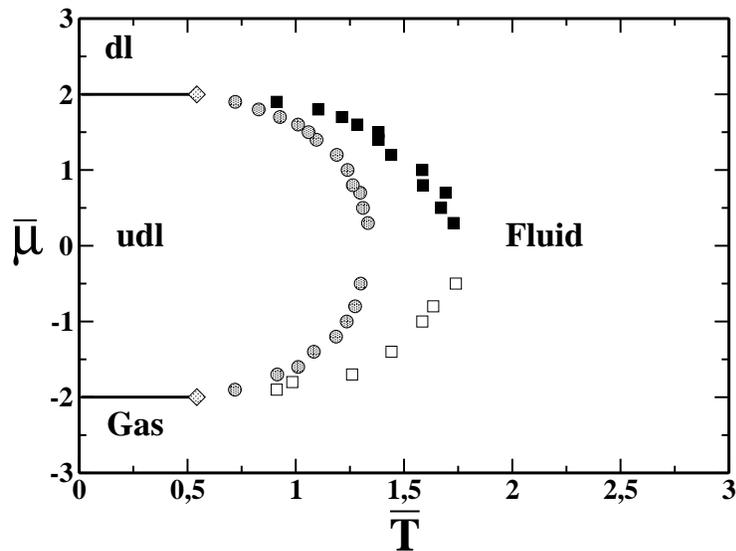}
\caption{$\bar{\mu} .vs.\bar{T}$  phase-diagram for the
lattice $20x20$. The solid lines are the first-order transitions, the 
circles are the critical line, the filled  squares are temperatures
of maximum density ($TMD$), the empty
squares are the temperatures of minimum density ($TmD$) and the diamonds are the tricritical points.}
\end{center}
\end{figure}

The $p.vs.T$ phase diagram (see  Fig.(13))  is constructed by
numerically integrating  simulations at constant pressure. It exhibits 
two liquid phases, a critical line and two tricritical points like
in Fig.(12). Close to the critical line, there is a line of 
temperature of maximum density  at constant pressure.

\vspace{1cm}
 
\begin{figure}
\begin{center}
\includegraphics[clip, scale=0.4]{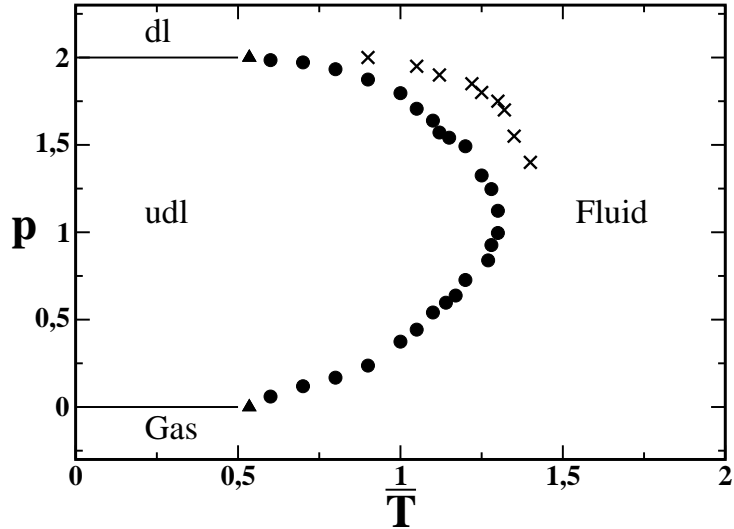}
\caption{ $p .vs. \bar{T}_0$ phase-diagram for the $20x20$ lattice. The solid lines are first-order transitions, the filled circles are 
the critical line, the crosses are the $TMD$ at constant pressure line and the filled triangles are the tricritical points.}
\end{center}
\end{figure}

From finite size
scaling analysis \cite{Ne99}, it is possible to 
estimate the critical temperature  of an infinite system by 
the following expression:
\begin{eqnarray}
\bar{T}_0(\bar{\mu})=T_{c}(\bar{\mu})\left(1+x_{0}L^{-1/2}\right)
\label{T0}
\end{eqnarray}
where $\bar{T}_0(\bar{\mu})$ is the critical  temperature of the 
finite  system, $T_{c}(\bar{\mu})$ is  the critical temperature 
of infinite system and $x_{0}$ a  parameter.

 Fig.(14) shows the critical 
temperatures of the finite systems
 as a function of the system size L. The value of $\bar{T}_0$
was  obtained from two different
methods: the maximum of the specific heat  at
a fixed chemical potential $\bar{\mu}=1.7$ and the minimum of the 
Binder cumulant at the same chemical potential. The 
extrapolated critical temperatures 
are  $\bar{T}_c(\bar{\mu})=0.92079$ 
and $\bar{T}_c(\bar{\mu})=0.92543$
respectively. 
The difference between these two values is within our error bars
of our simulations.

\begin{figure}
\begin{center}
\includegraphics[clip, scale=0.4]{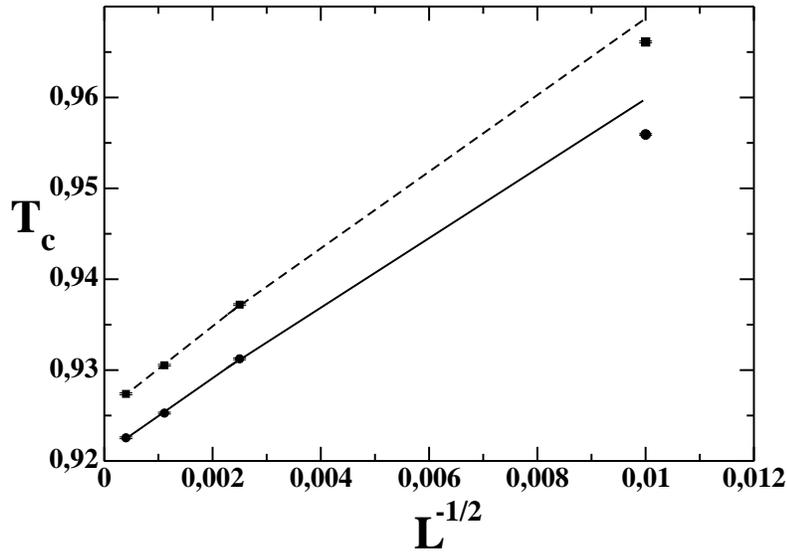}
\caption{ Minimum of the Binder cumulant (dashed line) and  maximum of the specific heat (solid line)  as a function of 
the system size $L=10,20,30,50$ for a fixed chemical potential $\bar{\mu}=1.7$.}
\end{center}
\end{figure}

\section{Conclusions}

In this paper the   phase-diagram of a two dimensional lattice
gas model with competing interactions 
was investigated by mean-field analysis and Monte
Carlo simulations. 
It was shown that this  system exhibits two liquid phases and a line
of density anomalies. 

The relation between the criticality  and the density anomaly 
in this model goes as follows.
The two liquid  and the gas phases  appear as a result of two competing
interactions: the  softened core that favors the formation 
of the  uniformly dilute liquid phase  and the chemical potential that
induces  the system to become completely filled or empty.  
For each soft-core interaction parameter, there is
a limit chemical potential beyond which the system is in
the dense liquid phase and another negative chemical potential 
beneath which the system is in the gas phase. The double
criticality arises from the competition between $\bar{\mu}$ and 
$V_1$.

In systems dominated by short-range attractive forces
the density increases on cooling.    
For  the soft-core potential studied here
similar behavior
is only observed when the short-range repulsion becomes
irrelevant ( high temperatures, $\bar{\mu}>2$ and $\bar{\mu}<-2$). For
$2>\bar{\mu}>-2$,  the soft-core
repulsion prevents the density to increase to one as the 
temperature is decreased.
Therefore, the same competition responsible for the appearance of
two liquid phases leads to the density anomaly. Similar analysis 
can be made when the pressure instead of the chemical potential
is kept constant.

 The presence 
of a  critical line instead of a single critical point as 
one could generally expect \cite{Po92} is not surprising.
Due to the lattice structure, the udl is not one single phase
but a region where two  different phases coexist: alternating empty/full rows and alternating empty/full columns. These two 
phases become critical together with the dense liquid phase 
at the tricritical point that is also the locus the where critical line
ends. The link between competing interactions and the presence of 
density anomaly and two liquid phases is being tested in other 
simple models
\cite{He04} \cite{Ol04}.

\vspace*{1.25cm}

\noindent \textbf{\large Acknowledgments}{\large \par}

\vspace*{0.5cm} This work was supported by the Brazilian science agencies CNPq,
FINEP, Capes  and Fapergs.

\end{document}